\documentclass{rspublic}

\newcommand{\be}{\begin{eqnarray}}
\newcommand{\ee}{\end{eqnarray}}

\newcommand{\vB}{{\bf B}}
\newcommand{\vE}{{\bf E}}
\newcommand{\vr}{{\bf r}}

\newcommand{\ffF}{{\sf F}}
\newcommand{\ffU}{{\sf v}}

\newcommand{\dxs}{{\rm d}x_s}
\newcommand{\ppt}[1]{\frac{\partial #1}{\partial t}}
\newcommand{\ppx}[1]{\frac{\partial #1}{\partial x}}
\newcommand{\binomial}[2]{\left( \!\! \begin{array}{c} {#1}\\{#2} \end{array} \!\! \right) }

\newcommand{\Grad}{\mbox{\boldmath $\nabla$}}
\newcommand{\Div}{\mbox{\boldmath $\nabla$} \cdot}

\newcommand{\Lap}{\nabla^2}

\newcommand{\myfig}[2]
{\centerline{\resizebox{!}{#1\textwidth}{\includegraphics{#2}}}}

\usepackage{graphicx}
\usepackage{amssymb}

\begin{document}

\title[Fields and self-force of spherical shell]{The fields and self-force of a constantly accelerating spherical shell}

\author{Andrew M. Steane}
%\ead{a.steane@physics.ox.ac.uk}
\affiliation{Department of Atomic and Laser Physics, Clarendon Laboratory, Parks Road, Oxford OX1 3PU, England.}

\label{firstpage}
\maketitle

\begin{abstract}{self-force, radiation reaction, hyperbolic motion}
We present a partial differential equation describing the electromagnetic potentials around a charge distribution undergoing
rigid motion at constant proper acceleration, and obtain a set of solutions to this equation. These solutions are
used to find the self-force exactly in a chosen case.
The electromagnetic self-force for a spherical shell of charge of proper radius $R$ undergoing rigid motion 
at constant proper acceleration $a_0$ is, to high order approximation,
$(2 e^2 a_0/R) \sum_{n=0}^\infty (a_0 R)^{2n} ((2n-1)(2n+1)^2(2n+3))^{-1}$,
and this is conjectured to be exact.
\end{abstract}

\section{Introduction}

The problem of self-force, and the related problem of radiation reaction, has a long history; see Rohrlich (1997) for a
brief review and Spohn (2004) for a longer survey. It is currently studied both because it is important in
the extreme physical conditions now realised in experiments (Bulanov {\em et al.} (2011)) and because it
continues to raise questions about the consistency and correct treatment of electromagnetism
(Gralla {\em et al.} (2009); Cremaschini and Tessarotto (2011)).

Lorentz (1904, 1909), Abraham (1905), Poincar\'e (1905, 1906) and Schott (1912) carried out pioneering work
on the understanding of self-force and performed approximate calculations for a rigid spherical shell whose centre
undergoes arbitrary motion. They obtained the lowest order terms 
in a power series solution, expanded in powers of the
radius $R$ of the sphere, its acceleration and derivatives thereof, and combinations of all these.
von Laue (1909) clarified and simplified the work of earlier authors by expressing it in a manifestly
Lorentz-covariant form, see eqn (\ref{fsmallR}).
Dirac (1938) adopted a different strategy, starting from a manifestly covariant Lagrangian, and reproduced
this equation; it is now commonly known
as the Lorentz-Dirac or Abraham-Lorentz-Dirac (ALD) equation. To be precise,
Dirac's version has a modified inertial term for reasons that we shall discuss below. 
The ALD equation includes terms of order $R^{-1}$ and $R^0$, and neglects terms of order $R$ and above.

Nodvik (1964) brought arguably greater rigour to the derivation using the Lagrangian formalism
and presented the power series expansion to $O(R^2)$.
Further work has clarified the interpretation of the force
and the associated energy-momentum movements in the field, 
and has extended the calculation of the self-force, such that
the `linear' contribution can be obtained exactly, in a beautifully neat and well-behaved expression
(eqn  (\ref{ffromv})). The word `linear' here does not mean, however, that all the contribution at
order $R^1$ is accounted for by this expression, since it neglects terms involving $R$ multiplied by
non-linear combinations of the velocity or its derivatives (see Caldirola (1956), Yaghjian (1992) and 
Rohrlich (1999, 2000) for details).

Throughout this century-long period, the study of self-force has
continued to raise doubts and questions concerning the correct interpretation or application of results. Many of these
issues were helpfully clarified by Yaghjian (1992) and Rohrlich (1965, 1990). For example, Yaghjian showed that
there is no `preacceleration' in the classical electromagnetic theory of small charged bodies; its 
supposed existence (discussed by authors too numerous to list) was based on a
misunderstanding of the approximations involved in the ALD equation. Numerous authors and textbooks
continue to assert that the ALD equation has pathalogical runaway solutions, but this too is an artifact
of its incorrect application, as we discuss in section \ref{s.discuss}, and as 
Gralla {\em et al.} (2009) and Rohrlich (2001), among others, have commented. A further, and
different, source of
confusion is the effect of Poincar\'e stresses in a body undergoing acceleration.

To have a concrete result with no approximation, for a non-trivial case, remains very helpful
both for pedagogy and as a test case for more general but approximate formulae.
Most previous work has been devoted to the case of arbitrary motion, endeavouring to extend the set of available
approximate results and understand them better. The present work takes the complimentary path of restricting
the motion to a simple case, and solving it exactly. We find the electromagnetic 
self-force for a spherical charged shell undergoing 
constant proper acceleration while maintaining constant proper size and shape. To our knowledge
this has not previously been done. The method of calculation involves a useful general observation about
the electromagnetic potentials in the case of arbitrary bodies undergoing rigid hyperbolic motion.

\section{Strategy}

The classical electromagnetic contribution to the self-force 
for an arbitrary distribution of charge, undergoing arbitrary motion, is
\be
{\bf f}_{\rm self} = \int  (\vE + {\bf v} \wedge \vB) \rho \, {\rm d}V
\ee
where $\rho$ is the volume density of charge, the integral is taken over all space (though of course it suffices to restrict it
to the region where $\rho \ne 0$) at some instant of time $t$ in a chosen inertial reference frame,
and the fields $\vE(t,x,y,z),\;\vB(t,x,y,z)$ are those produced at each point by the set of charges in question (owing to
their previous behaviour at some set of source events). Note that if $\rho$ is everywhere finite, then the fields
are also finite and continuous, and furthermore the contribution to the field at a given point $(x,y,z)$ made by the charge within a distance
$d$ of that point vanishes in the limit $d \rightarrow 0$ (because the quantity of charge scales as $\rho d^3$ but the fields
of a point charge scale as $1/d^2$). Hence no ambiguity arises about whether the integral above correctly excludes
the non-physical idea of a source interacting with itself: only pairwise interactions should be included, and only pairwise interactions
contribute to the integral.

In the case of a distribution of charge that is rigid, there must be further forces acting (for example, the Poincar\'e
stresses) and these further forces may be deemed `internal' and yet have a non-vanishing sum over the body.
Hence the total self-force includes both the electromagnetic part that is the main subject of this paper, and
a further part that we shall discuss in section \ref{s.discuss}.
The non-vanishing contribution from the `internal' forces does not break Newton's third law but rather is required
by it, since there is a continous transfer of momentum between the shell and the surrounding electromagnetic
field, mediated via the electric charge.

In the case of a spherical shell of charge, we treat a shell of finite thickness $w$, so that the volume density of charge is finite,
and then explore the limit $w \rightarrow 0$, such that the surface charge density $\sigma = \rho w$ remains finite.
We shall consider a shell which moves such that there exists a frame in which
the whole shell is at rest at some moment, so the ${\bf v} \wedge \vB$ term vanishes from the force. Let $R$ be the inner
radius of the shell in such a frame (the shell being spherical in that frame), and assume
the charge density $\rho$ is uniform between the inner and outer surfaces at $r=R$ and $r=R+w$.
For a sufficiently thin shell,
the field within the material of the shell (where $\rho \ne 0$) can be approximated by the linear form
\be
\vE(r) = \vE(R) + \frac{s}{w} \left( \vE(R+w) - \vE(R) \right)
\ee
where $\vE(r,\theta,\phi)$ is the field at radius $r$ from the centre of the sphere,  $s = r-R$ is the distance through the shell,
and we suppressed explicit indication of the dependence of the fields on polar angles $\theta,\phi$. Hence
\be
{\bf f}_{\rm self} = \int \int \rho \frac{1}{w}((w-s)\vE(R) + s \vE(R+w) ) {\rm d}s {\rm d}S
\ee
where ${\rm d}S$ is an element of surface.
Carrying out the integration over $s$ gives, in the limit $w \rightarrow 0$,
\be
{\bf f}_{\rm self} = \int \sigma \frac{\vE_- + \vE_+}{2}  {\rm d}S
\ee
where $\vE_- = \lim_{\epsilon \rightarrow 0} \vE(R-|\epsilon|)$ is the field on the interior surface of the shell, and
$\vE_+ = \lim_{\epsilon \rightarrow 0} \vE(R+|\epsilon|)$ is the field on the exterior surface of the shell. In other words,
to calculate the self-force it suffices to use the average of these two fields. This deals with the discontinuity in $\vE$ in the limit
$w \rightarrow 0$. Now, we know from Gauss' law what the size and direction of this discontinuity is: its size is $\sigma/\epsilon_0$ and
its direction is normal to the surface. For a spherically symmetric case, $\sigma$ is the same everywhere over the shell and we have
\be
\vE_+ = \vE_- + \frac{\sigma}{\epsilon_0} \hat{\vr}.
\ee
Since the contribution from $\sigma \hat{\vr}$ is radial and spherically symmetric, its integral gives zero contribution to the self-force.
Therefore we can also write
\be
{\bf f}_{\rm self} = \sigma  \int \vE_-   {\rm d}S =  \sigma  \int \vE_+  {\rm d}S.            \label{fselfmp}
\ee
In other words, one can use either the field on the interior surface or the field on the exterior surface to calculate the self-force.
For convenience, we will use the field on the interior surface.

\section{Simple calculation of the lowest order approximation}

Consider a particle undergoing hyperbolic motion in the $x$ direction, so that relative to some frame its position as a function of time is $y_s=z_s=0$ and
\be
x_s(t) = \sqrt{L^2 + t^2}   \label{xsLt}
\ee
where
\be
L = 1/a_0
\ee
is a natural distance scale set by the proper acceleration $a_0$.
The potentials and fields of a point charge $q$ undergoing such motion
are easily calculated, see for example Fulton \& Rohrlich (1960),
Eriksen \& Gr\/on (2000), Steane (2012).
Adopting rectangular coordinates and $c=1$, one finds the scalar and vector potentials (Li\'enard-Wiechart potentials)
\be
\phi = Q \frac{x \delta - t \zeta}{\zeta(x^2 - t^2)}, \;\;\; A_x = Q \frac{t \delta - x \zeta}{\zeta(x^2 - t^2)}, \; A_y = A_z = 0,   \label{phiAxAyAz}
\ee
where $Q = q/4 \pi \epsilon_0$ and
\be
\delta = L^2 + x^2 + y^2+z^2 - t^2, \;\;\; \zeta = \sqrt{\delta^2 - 4 L^2(x^2 - t^2) }.
\ee
In the following we will only need the solution at $t=0$, which is the moment when the particle is at rest in the frame under consideration.
At this moment one finds the magnetic field is $\vB=0$ and the electric field is
(adopting now cylindrical coordinates)
\be
E_x = -4 Q L^2(L^2 + \rho^2 - x^2)/\zeta^3, \;\;\; E_\rho = 8 Q L^2 \rho x/\zeta^3.  \label{Ex1}
\ee

Now consider a spherical shell of charge undergoing rigid hyperbolic motion. By {\em rigid} motion we mean motion such that, at any moment,
there is an instantaneous rest frame in which all parts of the body are at rest, and the physical dimensions of the body in each such frame
are constant (i.e. the same in all successive instantaneous rest frames), see for example Steane (2012). Such rigid motion has the property that
once the initial conditions and the worldline of one particle in the body are given, the worldlines of all other particles are fixed. For the
case of constant proper acceleration (i.e. hyperbolic motion) of one part of a body, the resulting constraint takes a particularly simple
form: every part of the body undergoes constant proper acceleration, but with a proper acceleration varying with distance from a plane
normal to the acceleration. Such a body is at rest relative to the well-known ``constantly accelerating reference frame"; the special plane is a horizon which can
conveniently be placed at $x=0$ by a suitable choice of origin. The worldline of each part of the shell is given again by  (\ref{xsLt}), but
now we must allow $L$ to be chosen for each part such that that part comes to rest at the correct position at $t=0$. This is easily done:
one sets $L$ equal to the location of the part at $t=0$. For example, for a spherical shell of proper radius $R$ centred at
$(L_0,0,0)$ one must use $L=L_0 + R \cos\theta$ where $\theta$ is the azimuthal angle relative to the centre of the shell and the $x$ axis.

\begin{figure}
\myfig{0.35}{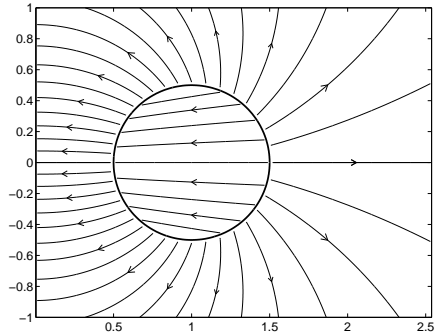}
\caption{Electric field lines in the $xy$ plane, in the instantaneous rest frame, for a spherical charged shell of 
radius $R=1/2$ undergoing rigid hyperbolic motion along the $x$ axis with proper acceleration $a_0 = 1/L_0 = 1$.}
\label{f.field}
\end{figure}

The observation of the previous paragraph makes it possible to write down an integral expression for the field at the centre
of the shell. A moment's thought about the geometry leads one to conclude that 
the integrand is obtained from Eq. (\ref{Ex1}) with the replacements $L\rightarrow x_s,\;  \rho^2 \rightarrow R^2-(x_s-L_0)^2, \;
x \rightarrow L_0$, leading to
\be
E_x(L_0,0,0) = \frac{Q}{4\pi R} \int_{L_0-R}^{L_0+R}  \int_0^{2\pi} {\rm d}\phi \dxs \frac{4 {x_s}^2 \left(2 L_0^2-R^2-2 L_0 x_s \right)}{R^3 \left(R^2+4 L_0 x_s \right)^{3/2} }
\ee
and $E_y = E_z = 0$. The integration is straightforward. One obtains
\be
E_x(L_0,0,0) = \frac{Q}{L_0^2}\left( -\frac{2}{3}\frac{L_0}{R} + \frac{2}{15} \frac{R}{L_0} \right).  \label{ExL000}
\ee

To find the self-force of such a shell, to lowest order approximation, is now extremely easy. One simply multiplies this expression by $q$:
\be
f_{\rm self} = -\frac{2}{3} \frac{e^2}{L_0 R}  +\frac{1}{L_0^2} O(R/L_0) =  -\frac{2}{3} \frac{e^2 a_0}{R}  +  O(a_0^3 R)
\label{fself0}
\ee
where to reduce clutter we introduced $e^2 \equiv q^2/(4\pi\epsilon_0)$.
To see that this expression is accurate to lowest order, as claimed, argue as follows. The self-force is in fact given by the
either of the integrals in (\ref{fselfmp}), of which we choose the first. The field at the interior surface, $\vE_-$, may be
obtained from a Taylor expansion of the field throughout the empty space
in the interior of the shell, expanded about the point $(L_0,0,0)$.
One finds that $\vE_-$ is given by Eq. (\ref{ExL000}) plus
further terms forming a power series in $(x_s-L_0)/L_0$. 
Using the symmetry of the charge distribution, the linear terms
cancel in the calculation of the self-force, and the next contribution is at order $(R/L_0)^2$ compared to the first.
In summary, the field is, to good approximation, uniform throughout the interior (c.f. figure
\ref{f.field}), and to find the self-force one simply multiplies
this uniform field by the total charge of the shell. Thus we obtain Eq. (\ref{fself0}). This
is the simplest derivation of it of which I am aware.

\section{The potentials}

We now turn to the task of calculating the electric field, and hence the self-force,
exactly. We will obtain the field from the potentials via $\vE = -\Grad \phi - \partial {\bf A}/{\partial t}$.
In empty space $\Div \vE=0$, hence
\be  
\Lap \phi = -\ppt{} \Div {\bf A} = - \frac{\partial^2 A_x}{\partial x \partial t}     \label{lapphi}
\ee
where we used (\ref{phiAxAyAz}) in the final step. This result is valid for the potentials
of a single constantly accelerating point charge, and since for a distribution of charge
the potentials simply add and the equation is linear, it remains valid for an arbitrary
distribution of point charges, each undergoing constant proper acceleration in the $x$ direction (not necessarily all having the same acceleration). Evaluating $\partial A_x/\partial t$ for a single point charge using (\ref{phiAxAyAz}), one finds at $t=0$,
\be
\left.\ppt{A_x} \right|_{t=0} = \frac{Q \delta}{\zeta x^2} = \frac{\phi}{x}.
\ee
It follows that the electric field can be obtained from
\be
\vE = -\Grad \phi - \frac{\phi}{x} \hat{\bf x}           \label{Efromphi}
\ee
and Eq (\ref{lapphi}) can be written
\be
x^2 \Lap \phi + x \ppx{\phi} - \phi = 0.  \label{PDE}
\ee
This is a linear homogeneous second order differential equation for the potential
in free space. It may be regarded as a replacement for the Laplace equation;
it applies in the free space
around a point charge undergoing hyperbolic motion in the $x$ direction.

Using the same argument concerning rigid motion as we presented
after Eq (\ref{Ex1}), we will now show that Eq (\ref{PDE}) has a wider range
of validity. Consider a distribution of charge undergoing rigid
hyperbolic motion in the $x$ direction. Then in the instantaneous rest
frame, the total scalar potential at any field point $(x,y,z)$ is given by
\be
\phi(x,y,z) = \int \int \int {\rm d}x_s {\rm d}y_s {\rm d} z_s
\rho(x_s,y_s,z_s)
\tilde{\phi}(x_s; \, x,\, y-y_x, \, z-z_s)        \label{phifromtilde}
\ee
where $\rho$ is the charge density and $\tilde{\phi}$ is the potential per
unit charge obtained from Eqn (\ref{phiAxAyAz}):
\be
\tilde{\phi}(L;\, x,\,y,\,z) = \frac{L^2 + x^2 + y^2+z^2}
{x [(L^2 + x^2 + y^2+z^2)^2 - 4 L^2 x^2 ]^{1/2}}.
\ee
Note that we are evaluating the potential quite correctly using the
distribution of charge at its `present' location $(t=0)$, not the location
it had at the source events, because we can take advantage of the fact that
we already know how to calculate what contribution each source event makes
to the potential `now' (at $t=0$) for the simple type of motion under consideration.

The important property of Eq (\ref{phifromtilde}) for our purposes is that
every source charge presents a contribution that is calculated by substituting
the {\em same} value for $x$ into the formula for $\tilde{\phi}$, and therefore
every contribution satisfies (\ref{PDE}) with a common value of $x$, and therefore
(\ref{PDE}) applies to the net resulting $\phi$ (the one given by (\ref{phifromtilde})).
Hence Eq (\ref{PDE}) describes the potential in the free space around, or
in cavities within, any distribution of charge that is
undergoing rigid hyperbolic motion.
(In obtaining it we have made a gauge choice that makes
$\phi \rightarrow 0$ at positions far from any finite charge contained in a finite
region.)

The problem of finding the potential inside an accelerating spherical shell
can now be addressed by seeking a potential function $\phi$ that satisfies
(\ref{PDE}) and that agrees with a suitable boundary condition.

\subsection{Solution of the differential equation}

In order to solve (\ref{PDE}), introduce another function $V(x,y,z) \equiv \phi/x$. Then one finds that $V$ satisfies
\be
x \Lap V + 3 \ppx{V} = 0       \label{PDEV}
\ee
If $V=V(x)$ (i.e. a solution with no dependence on $y$ and $z$) this leads to a Cauchy-type equation with
general solution
\be
V(x) = a_0 + a_1 \frac{1}{x^2}
\ee
More generally, the equation is not readily separable, but we can look for solutions of a form
suitable to the type of problem we are treating. We would like solutions which are non-singular at
$\rho=0$ and which have axial symmetry, so we try the ansatz
\be
V(x,\rho) = \sum_{n=0}^N f_n(x) \rho^{2(N-n)}       \label{Vansatz}
\ee
where $\rho = (y^2 + z^2)^{1/2}$ is the radial coordinate in a cylindrical coordinate system.
Substituting this ansatz into (\ref{PDEV}) and using $\Lap \rho^n = n^2 \rho^{n-2}$, one finds
\be
\sum_{n=0}^N \left( x f_n'' + 3 f_n' \right) \rho^{2(N-n)} + 4(N-n)^2 x \rho^{2(N-n-1)} f_n = 0.  \label{sumdff}
\ee
Since this is to be satisfied at all $\rho$, we can set the coefficient of $\rho^{2(N-k-1)}$
equal to zero:
\be
x f_{k+1}'' + 3 f_{k+1}' + 4(N-k)^2 x f_k = 0.
\ee
Integrating, we have
\be
x f_{k+1}' + 2 f_{k+1} = -4(N-k)^2 \int x f_k {\rm d}x  \label{fkp}
\ee
which applies for values of $k$ between $0$ and $N-1$. The coefficient of $\rho^{2N}$ in (\ref{sumdff})
gives
\be
x f_{0}'' + 3 f_{0}' = 0.
\ee
This has the general solution $f_0 = a_0 + a_1/x^2$, which can be substituted into
the right hand side of (\ref{fkp}) in order to find $f_1$, and hence $f_2$ and so on.
Table \ref{t.sol1} shows the first few solutions found this way.

%\begin{table}
%\begin{eqnarray*}
%&& -\rho^4/ x^2 + 3 x^2 + (8 \rho^2  - 4 x^2) \log  x, \\ 
%&& 3 \rho^4 - 6 \rho^2 x^2 + x^4,\\ 
%&& -\rho^2/ x^2 + 2 \log  x,\\
%&&  2 \rho^2 - x^2, \\
%&& -1/x^2, \\
%&& 1
%\end{eqnarray*}
%\end{table}

\begin{table}
\begin{eqnarray*}
\begin{array}{lcl}
1, && 1/x^2, \\
2 \rho^2 - x^2, && \rho^2/ x^2 - 2 \log  x,\\
3 \rho^4 - 6 \rho^2 x^2 + x^4, && \rho^4/ x^2 - 3 x^2 - 4(2 \rho^2  - x^2) \log  x, \\ 
4 \rho^6 - 18 \rho^4 x^2 + 12 \rho^2 x^4 - x^6, &&
\rho^6/x^2 - 27 \rho^2 x^2 + 7 x^4 - 6(3 \rho^4 - 6 \rho^2 x^2  + x^4) \log x
\end{array}
\end{eqnarray*}
\caption{The first few solutions of Eq (\protect\ref{PDEV}) having the form
(\protect\ref{Vansatz}).}
\label{t.sol1}
\end{table}

In the rest of this paper we will restrict attention to solutions having the form
\be
f_k(x) = \sum_{m=0}^k c^{k}_m x^{2m}       \label{fkx}
\ee
where the $c^k_m$ are constants and $k\ge 0$. This restriction is merely for convenience: one
can expand  $V(x,\rho)$ for any given charge distribution in terms of any sufficiently complete set
of basis solutions, and this is one such set for our problem.
Substituting (\ref{fkx}) into (\ref{fkp}) we obtain
\be
\sum_{m=0}^{k+1} 2(m+1) c^{k+1}_m x^{2m} = 
-4(N-k)^2 \sum_{m=0}^k c_m^k \frac{x^{2(m+1)}}{2(m+1)}
\ee
Hence
\be
c^{k+1}_{m+1} = \frac{-(N-k)^2}{(m+1)(m+2)} c^k_m
\ee
and $c^k_0$ are undetermined constants to be found from boundary conditions. For
example, with $N=2$ one finds $c^0_0$ is undetermined, then $c^1_1 = -2 c^0_0$ and
$c^1_0$ is undetermined. These are in turn used to find $c^2_2 = (-1/6)c^1_1=c^0_0/3$
and $c^2_1 = (-1/2) c^1_0$. Finally, $c^2_0$ is undetermined. 
For any given value of $N$ we thus find a solution involving a 
set of $N+1$ undetermined constants, each of which multiplies a polynomial function of
$x$ and $\rho$. The polynomial function which appears with the factor $c^n_0$ is the one involving
$\{c^n_0, \; c^{n+1}_1, \;c^{n+2}_2,\; \ldots ,\; c^N_{N-n} \}$. To be precise, it is
$\sum_{m=0}^{N-n} c^{(n+m)}_m x^{2m} \rho^{2(N-n-m)} $.
By using the above recurrence relation, the terms in the sum are easily found, and one obtains
\be
V_n(x,\rho)  = \sum_{m=0}^n \frac{ (-1)^m }{m+1} \binomial{n}{m}^{\! 2} x^{2m} \rho^{2(n-m)}
= \rho^{2n} {_2 \! F_1} \! \left(-n,-n,2,-x^2/\rho^2 \right).  \label{Vnxrho}
\ee
where $(\stackrel{n}{\mbox{\scriptsize $m$}})=n!/n!(n-m)!$ is a binomial coefficient and 
$ {_2 \! F_1}$ is the hypergeometric function. For $n=0 \ldots \infty$ we thus obtain an infinite set 
of linearly independent solutions of Eqn (\ref{PDEV}). The first few of these functions are shown
in the left hand column of table \ref{t.sol1}, in each case multiplied by $(n+1)$ to make the coefficients
all integer.

The electric field given by these solutions is (using Eqn (\ref{Efromphi}) and
$\phi = x V_n$)
\be
E_{x,n} &=& -x \ppx{V_n}-2 V_n = 2 \sum_{m=0}^n (-1)^{m+1} \binomial{n}{m}^{\! 2} x^{2m} \rho^{2(n-m)} \\
&=& -2 \rho^{2n} {_2 \! F_1}\! \left(-n,-n,1,-x^2/\rho^2 \right),        \label{Efieldn}
\ee
and
\be
E_{\rho,n} &=&  -x \frac{\partial V_n}{\partial \rho} = 2 \sum_{m=0}^{n-1} (-1)^{m+1} \frac{n-m}{m+1} \binomial{n}{m}^{\! 2} x^{2m+1} \rho^{2(n-m-1/2)} \\
&=& -2 n x \rho^{2n-1} {_2 \! F_1}\! \left(1-n,-n,2,-x^2/\rho^2 \right).    \label{Efieldrhon}
\ee

\section{Interior field and self-force of the accelerating spherical shell}

We now apply the ideas of the previous section to the case of the spherical
shell of charge undergoing rigid hyperbolic motion. Without loss of generality, choose
distance units such that the centre of the shell comes to rest at $x=1$ (this amounts
to setting $L_0 = 1$ in Eqn (\ref{fself0})). Then the surface of the shell is at $\rho^2 = R^2 - (x-1)^2$.
By substituting this into (\ref{Efieldn}) one finds the contribution to the electric field at the
(interior surface of the) shell for
each of the potential solutions identified in the previous section. For a uniformly charged spherical
shell with unit total charge, the contribution to the self-force is then given by
\be
f_{{\rm self},n} &=& \frac{1}{2 R} \int_{1-R}^{1+R} E_{x,n}(x,\rho(x)) {\rm d}x = 2 \sum_{m=0}^n \frac{(-1)^{m+n+1}}{2m+1} R^{2m}
\binomial{n}{m} \\
&=& 2 (-1)^{n+1}  {_2 \! F_1}\! \left(1/2,-n,3/2, R^2 \right).
\ee
For example, the first few values of this expression are given by
\be
\frac{1}{2} f_{{\rm self},n} = -1, \;
 \left(1-\frac{R^2}{3} \right), \;
 \left( -1 + \frac{2R^2}{3} - \frac{R^4}{5} \right), \; 
 \left(1-\frac{3 R^2}{3} + \frac{3R^2}{5} - \frac{R^6}{7} \right),\; \ldots
\ee

\begin{figure}
\myfig{0.35}{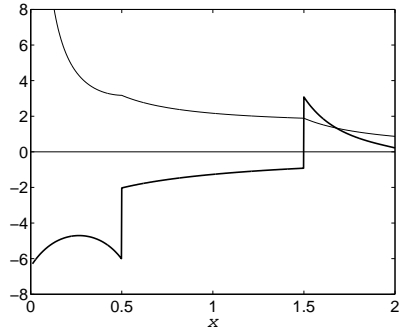}
\caption{Electric field $E$ (thick line) and potential $\phi$ (thin line) on the $x$ axis for a sphere of radius $R=1/2$ with unit charge
accelerating in the $x$ direction with proper acceleration $a_0 = 1/L_0 = 1$.}
\label{f.phiE}
\end{figure}

To find the self-force of the shell, it remains to identify what linear combination of these solutions matches the potential, and hence the field, in
the interior of a uniformly charged shell undergoing rigid hyperbolic motion. To find this we first evaluate the potential on the axis of
the shell, by performing the integral in (\ref{phifromtilde}) for $y=z=0$:
\be
\phi(x,0,0) &=& 
\frac{1}{2 R} \int_{1-R}^{1+R}
\frac{ R^2 + 2 x_s + x^2 -1}{x \sqrt{(R^2 + 2x_s + x^2-1)^2 - 4 x_s^2 x^2}} {\rm d}x_s \\
&=&
\frac{|\beta_-| - |\beta_+|}{4 R s^2 x} + \frac{i x (R^2+s^2)}{4 R s^3} \log
\left[ \frac{s|\beta_+| + i(\beta_+ -2Rx^2)}{s|\beta_-| + i(\beta_- +2Rx^2)} \right]
\ee
where $s = \sqrt{x^2-1}$ and $\beta_{\pm} = (1\pm R)^2-x^2$. This function is plotted  in figure \ref{f.phiE} (note, the function
is real since the argument of the log has unit modulus when $s$ is real).
For $1-R < x < 1+R$ (that is, for points inside the sphere), it takes the form
\be
\phi(x,0,0) = 
\frac{-R^2 + x^2-1}{2Rx(x^2-1)} + \frac{x(R^2 + x^2-1)}{4R(x^2-1)^{3/2}} \tan^{-1}\left[\frac{2\sqrt{x^2-1}}{2 - x^2}\right]
\label{phix00}
\ee
We will now find the potential throughout the interior of the spherical shell by finding
a function that satisfies (\ref{PDEV}) and that matches the known potential and all
its derivatives with respect to $x$ at the point $(x,y,z)=(1,0,0)$ (the centre of the sphere).
This may be compared with matching a boundary condition along an infinite line, since for
an analytic function, the value of the
function and all its derivatives at a point suffice to define the function
along a line by Taylor expansion. Of course the potential given by the shell actually has
a discontinuous change in slope at $x=1 \pm R$, but this is irrelevant to the method.
By using $\phi$ and all its derivatives at one point we gain sufficient information to
fix the potential throughout the interior of the shell.

We employ the solution
\be 
V(x,\rho) = \sum_{n=0}^N a_n V_n(x,\rho)           \label{VN}
\ee
where $V_n(x,\rho)$ is given by (\ref{Vnxrho}), the constants $a_n$ are to be determined,
and the solution becomes exact in the limit $N \rightarrow \infty$.
We find the constants by writing down the Taylor expansion of both $V(x,0)$ and
$\phi(x,0,0)$ about the point $x=1$, and equating coefficients of powers of $(x-1)$.
This gives a set of $N+1$ simultaneous equations for the $a_n$ which can be solved by matrix
inversion. For example, for $N=2$ one finds
\be
a_0 = \frac{34}{15 R} + \frac{122}{105}R,\;
a_1 = \frac{56}{15 R} + \frac{88}{35}R,\;
a_2 = \frac{9}{5R} + \frac{9}{7}R .
\ee
The self-force is then given by
\be
f_{\rm self} \simeq \sum_{n=0}^N a_n f_{{\rm self},n}              \label{fselfsum}
\ee
where the expression becomes exact in the limit $N \rightarrow \infty$ and
for given finite $N$ one obtains an exact expression for 
the terms up to order $R^{2N-3}$ (see appendix). The above values for $a_0,a_1,a_2$ when $N=2$ yield,
for example,
\be
f_{\rm self} \simeq -\frac{2}{3R} + \frac{2}{45}R - \frac{358}{525}R^3 - \frac{18}{35} R^5  \label{fsol2}
\ee
in which the first two terms are exact. Further details of the Taylor expansions and
the solution coefficients are given in the appendix. The end result is that the 
self-force on the uniformly charged spherical shell undergoing rigid hyperbolic motion is
\be
f_{\rm self} = \frac{2 e^2}{R L_0} \sum_{n=0}^\infty \frac{(R/L_0)^{2n}}
{(2n-1)(2n+1)^2(2n+3)},   \label{fselfexact}
\ee
where I have calculated this result for terms up to order $R^{101}$, and I conjecture
its validity at all orders. This is the main result of this paper.
At $L_0=1$ and $e^2=1$ the first few terms in this series are
\be
f_{\rm self} \simeq -\frac{2}{3 R} + \frac{2}{45}R + \frac{2}{525} R^3 +
\frac{2}{2205} R^5 + \frac{2}{6237} R^7 +\cdots .   \label{fseries}
\ee
It is remarkable that the result takes such a simple form. One expects a power series in
odd powers of $R$, but it is striking that the coefficients have such a simple expression. The
coefficients in the Taylor series for $\phi(x,0,0)$ are, by contrast, much more complicated.

The lowest order terms in the expression (\ref{fseries})
for $f_{\rm self}$ can be found by hand. The series expansion and matrix inversion
method described above is essentially simple if laborious; it can easily be executed
in a computer algebra system such as {\em Mathematica}. This is how the series up
to order $R^{101}$ was calculated. The problem of finding the expression
for an arbitrary coefficient in the series, and thus confirming the (eminently reasonable)
conjecture that the result is valid at all orders, is more difficult and remains open.
As a further check, the field inside the shell was calculated by direct numerical
integration of the fields of a point charge given in Eqn (\ref{Ex1}), and compared
with the one obtained from equations (\ref{Efieldn}) and (\ref{Efieldrhon}),
multiplied by the coefficients $a_n$ and summed. This confirms, to within numerical precision,
that one obtains the field correctly at points away from the $x$ axis (as well as on it).

\begin{figure}
\myfig{0.4}{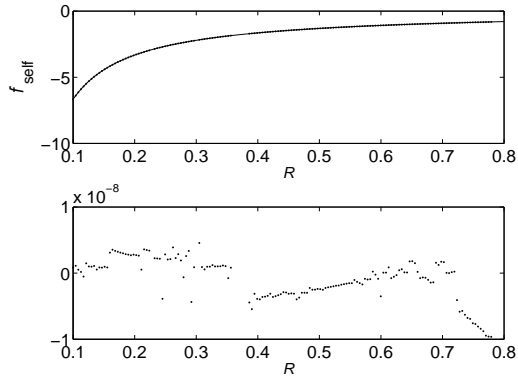}
\caption{Top: self-force as a function of $R$, for $L_0=1$, as calculated by numerical integration (points) and
by eqn (\ref{fselfexact}) (line). Bottom: the difference between the numerical and analytical results.}
\label{f.force}
\end{figure}

Numerical integration was also used to check the overall result (\ref{fselfexact}).
The field at the interior surface of the shell was obtained
by calculating the field (numerically)
at four points near the surface and then extrapolating to the surface
using a cubic fit; this enables the singular behaviour to be avoided. The
numerically-calculated self-force is compared with the prediction
(\ref{fselfexact}) in figure \ref{f.force}. By fitting a polynomial in $R^2$ to the numerical
calculation of $R f_{\rm self}$, the first few coefficients in (\ref{fselfexact}) were confirmed to good
approximation (one part in $10^{9,6,3}$ for the term of order $R^{-1,1,3}$ respectively). However the numerical
calculation rapidly runs out of precision in obtaining the higher order terms by this
`brute force' method, whereas the analytical result is easy to obtain for many more orders.

\section{Discussion}  \label{s.discuss}

Equation (\ref{fselfexact}) is the result of a fully relativistic calculation, and
is believed to be exact. It gives the spatial part of the self-four-force in the
instantaneous rest frame. The temporal part is zero in that frame. Since the resulting
four-vector is parallel to the four-acceleration, it may be written
\be
\ffF_{\rm self} = \frac{2 e^2}{R} \dot{\ffU} \sum_{n=0}^{\infty}
\frac{ (R a_0)^{2n}} {(2n-1)(2n+1)^2(2n+3)}              \label{fourself}
\ee
where $\ffU$ is the four-velocity (in an index-free notation) and the dot signifies the derivative with respect to proper time. The equation of motion of the spherical shell of charge is
\be
\ffF_{\rm other} + \ffF_{\rm self} = m_0 \dot{\ffU}           \label{eqmotion}
\ee
where $\ffF_{\rm other}$ is the four-force applied to the shell by some other influence, which must, in this case, take such
a form that, when combined with $\ffF_{\rm self}$, it gives to the shell the motion that has been assumed. $m_0$ is the
`bare' rest mass of the shell---that is, the rest mass which enters on the right hand
side of the equation of motion {\em of the material of the shell itself}. This
does not include a contribution from the energy-momentum density of the electromagnetic field
sourced by the charge of this shell, which is quite correct. To find out about the dynamics of the field sourced by the shell,
one can appeal to the conservation of energy-momentum which is perfectly obeyed
in classical electromagnetism. The force {\em from} the field {\em on} the charge (i.e. the electromagnetic 
self-force $f_{\rm self}$)
implies the presence of an equal and opposite `force on' (i.e. rate of injection of momentum into) the 
electromagnetic field. Hence the
field sourced by the spherical shell must be acquiring energy-momentum at the rate $-\ffF_{\rm self}$.

Now, we assumed that the motion of the shell was rigid. Therefore, the 4-force $\ffF_{\rm other}$ must
both oppose the tendency of the electromagnetic force to explode or otherwise deform the proper shape of
the shell, and it must also provide the correct amount of force to result in the assumed net acceleration of
the shell. It makes sense, therefore, to write
\be
\ffF_{\rm other} = \ffF_{\rm ext} + \ffF_{\rm P}
\ee
where $\ffF_{\rm ext}$ is provided by an external force-producing entity such as a rod pushing
on the sphere, and $\ffF_{\rm P}$ is deemed `internal', it may be regarded 
as the integral of the Poincar\'e stresses. 

To clarify the separation into $\ffF_{\rm ext},\; \ffF_{\rm P}$ and
$\ffF_{\rm self}$ it helps to imagine a deliberately artificial scenario. Consider a non-rigid latex balloon
with a uniform coating of charge attached to it by glue, and suppose that
inside the balloon there is a programable device which can push out or pull in the interior surface of the balloon
(the latex itself having negligible tension).
We plan to push this balloon from the outside using a wooden rod, so as to cause hyperbolic motion of its centre.
Before doing so, we first program the device to produce, starting from some preset starting time, whatever forces
are calculated to maintain rigid motion of the balloon, given that it is going to be pushed externally by
our rod. The internal device cannot, of course, produce any arbitrary combination of forces but
must respect Newton's
third law. Therefore we shall have to plan for the external force to have a suitable form in order to obtain
a self-consistent solution. Once this is done, we can launch the experiment. Then the three terms 
$\ffF_{\rm ext},\; \ffF_{\rm P}, \;\ffF_{\rm self}$ 
are provided by the rod, the internal device and the electromagnetic field sourced by the balloon, respectively.

In this separation of forces, $\ffF_{\rm P}$ may or may not be called a self-force. 
From the point of view of the surface of the balloon, $\ffF_{\rm P}$ is
an `external' force, i.e. one not sourced by the rubber of the balloon's surface, nor by the charge and
its associated electromagnetic field. However, from the point of view of an outside observer observing
the sphere has a whole, $\ffF_{\rm P}$ is a self-force, because it contributes something in addition
to the force $\ffF_{\rm ext}$ that `the rest of the world' is providing.

The assumed motion can also be obtained to good approximation in a less artificial scenario.
For a very stiff sphere, for example one made of diamond and given a uniform coating of charge, $\ffF_{\rm ext}$
could, for example, be the force from a uniform external electric field applied in the $x$ direction. Such an applied
field will tend to stretch the sphere (since it gives to all parts of the sphere the same acceleration in the instantaneous
rest frame, whereas we want the part momentarily at rest at given $x$ to have acceleration proportional to $1/x$).
The self-force is somewhat larger at the back of the sphere, and therefore also tends to stretch the sphere. If the stiffness of
the sphere is sufficient to make this stretching negligible, then rigid hyperbolic motion will result in a uniform applied electric field.
Another way to realize the conditions approximately is to hold the sphere fixed in a nearly-uniform gravitational field.

In the limit of a small sphere, one expects to obtain the ALD equation
\be
\ffF_{\rm self} \simeq  \frac{2}{3}e^2\left( -\frac{\dot{\ffU}}{R} + \ddot{\ffU} - \dot{\ffU}^2 \ffU \right) 
\label{fsmallR}
\ee
where we use the metric signature ${\rm Tr}(g)=+2$ so $\dot{\ffU}^2 = a_0^2$.
The three terms on the right hand side are called the inertial, Schott and loss (or radiation) terms, associated
respectively with the provision of four-momentum to the field bound to the charge, its redistribution within the
bound field, and the provision of four-momentum to the radiated field.
Hyperbolic motion is characterized by $\ddot{\ffU} = \dot{\ffU}^2 \ffU$ so exhibits a well-known balance
between the last two terms. Hence we expect $\ffF_{\rm self} \simeq -(2/3) e^2 \dot{\ffU}/R$, which is
the first term (the one given by $n=0$) in our series (\ref{fselfexact}). The next term in the series is
of order $R a_0^3$ and is not accounted for in the approximations leading to Eqn (\ref{fsmallR}).

{\em An aside on point-like particles}.
For the avoidance of all confusion, it should be emphasized that the ALD equation is quite correct,
and when used in the equation of motion (\ref{eqmotion}) it gives correct solutions with no pathalogical
behaviour, within the approximations
it assumes. Those approximations include that the applied force should not vary significantly on the timescale
$2R$ (the time for light to cross the sphere)---this gets rid of the `pre-acceleration' phenomenon
(see Yaghjian (1992)).
One must also insist that $m_0 \ge 0$, of course---this gets rid of the runaway solutions.
The introduction of negative mass into classical physics would produce pathalogical predictions
in all sorts of scenarios; the fact that it would do so here has nothing particularly to do with
electromagnetism. However, what one can correctly infer is that the ALD equation only
describes a small spherical shell if the `observed mass', defined by $m = m_0 + 2 e^2/3R$, is larger
than $2e^2/3R$.
One cannot by this route obtain an equation of motion for a pointlike particle with finite
charge and finite observed mass. In classical physics such an entity is simply impossible
(it has infinite field energy and infinitely negative bare mass); in quantum physics it is also
impossible, in the sense that a delta-function-like wavefunction is never achieved in practice.
Gralle {\em et al.} (2009) have shown how to obtain the equation of motion, in classical 
electromagnetism, for a small object whose total mass and charge tend to zero in a suitably
smooth way as the size of the object tends to zero. They achieve this by asserting energy-momentum
conservation for the total stress tensor, thus making the Lorentz force equation a derived not
an axiomatic part of the theory (c.f. Steane (2012) section 16.4). The ALD equation then emerges as
the lowest-order correction to the force on a finite-sized body; see their paper for details.
The connection to the present work is
twofold. First, both here and in Gralle {\em et al.} the calculation avoids
unphysical (and mathematically dubious) properties such as infinite field energy. Secondly,
their method should reproduce our result to all orders; this would be a useful further check
(to all orders in $R$ if possible, or failing that then to a few orders).

Another important approximate expression for the self-force of a spherical shell
undergoing arbitrary rigid  motion, (that is, relativistically rigid motion in which
the proper acceleration may vary) is (Caldirola (1956), Yaghjian (1992))
\be
\ffF_{\rm self} \simeq  \frac{2}{3}\frac{e^2}{R}\left( \frac{\ffU(\tau-\tau_R)}{\tau_R} + \frac{\ffU(\tau) \cdot \ffU(\tau-\tau_R)}{\tau_R} \ffU(\tau)  \right)             \label{ffromv}
\ee
where $\ffU(\tau)$ is the four-velocity at proper time $\tau$ and $\tau_R = 2R$ is the
time taken for light to cross the sphere. This result is more accurate than (\ref{fsmallR}) in that it
evaluates a Taylor expansion for the field to all orders in the derivatives of $\ffU$, but it still makes a linear
approximation in which all nonlinear powers of the acceleration and its derivatives are neglected.
In the instantaneous rest frame, the spatial part of this equation gives the three-force
\be
{\bf f}_{\rm self} \simeq  \frac{2}{3}\frac{e^2}{2 R^2} {\bf v}(\tau-\tau_R) .
\ee
which for hyperbolic motion ($v(\tau) = \tanh(a_0\tau)$) leads to 
\be
f_{\rm self} \simeq  e^2 \left( -\frac{2}{3}\frac{a_0}{R} + \frac{8}{9} R a_0^3 + \cdots \right).
\ee
Comparing this with (\ref{fselfexact}), we find that for hyperbolic motion
Eqn (\ref{ffromv}) correctly gives the inertial term but 
it should not be expected to, and does not, correctly give the term of order
$R a_0^3$, which is the leading term after the inertial term is accounted for (in the special case
of hyperbolic motion).

In an important and extensive paper, Nodvik (1964) gave a treatment of self-force in electromagnetism
using the calculus of variations to obtain covariant Euler-Lagrange equations for a body undergoing
relativistically rigid motion. 
He then obtains a perturbative expansion for the self-force up to order $R^2$ in our terminology. 
His general expression (eqns (7.21)-(7.24) of Nodvik (1964)) allows for rotation and an arbitrary 
distribution of charge. In the case of a non-rotating spherical shell it yields
\be
\ffF_{\rm self}^{\rm Nodvik}
&=& \frac{2}{3} e^2 \left\{
 -\frac{3}{4} \frac{\dot{\ffU}}{R}
+ \ddot{\ffU} - \dot{\ffU}^2 \ffU 
+ R \left[ 2(\dot{\ffU} \cdot \ddot{\ffU}) \ffU + \dot{\ffU}^2 \dot{\ffU} - \frac{2}{3} \dddot{\ffU} \right] \right. 
\nonumber \\
\lefteqn{
+ \left. \frac{R^2}{3} \left[ \ffU^{(4)} + (\ffU \cdot \ffU^{(4)}) \ffU - 2 \dot{\ffU}^2( \ddot{\ffU} - \dot{\ffU}^2 \ffU )
- 6 ( \dot{\ffU} \cdot \ddot{\ffU} ) \dot{\ffU} \right]  \right\} 
+ O(R^3)}
\ee
where $\ffU^{(n)} = {\rm d}^n \ffU/{\rm d} \tau^n$.
Applied to the case of constant proper acceleration, and setting $e^2=a_0=1$, this yields
\be
\ffF_{\rm self}^{\rm Nodvik} = -\frac{1}{2R} + \frac{2}{9} R + O(R^3),  \label{Nodvik}
\ee
which should be compared to our eqn (\ref{fseries}). The two results are in agreement for the even powers of
$R$, but not for the odd powers. This is because eqn (\ref{fseries}) gives the contribution purely from the
electromagnetic field sourced by the shell, whereas (\ref{Nodvik}) is the sum of this and the contribution
from Poincar\'e stresses. This fact is, however, less than clear in Nodvik's presentation which lumps the
self-electromagnetic
and the other part together. The same issue was already noted by Dirac (1938) and is exhibited, for example, in the 
lowest order term (the inertial term). Since the inertial term can be lumped together with the mass term in the
equation of motion and called `renormalisation', this difference has no easily observable consequence. However,
this should not be taken to mean that there is any doubt about the calculation. The electric field predicted by
Maxwell's equations is certainly the one we have calculated, and there is no need to abandon the Lorentz force
equation when calculating the force on a continuous distribution of charge; it is the simplest equation
consistent with energy-momentum conservation. Therefore eqn (\ref{fself0}) is also correct.
In the method adopted first by Dirac and then by Nodvik and many later workers, the rigidity of the charge
distribution is accounted for by including it as a constraint in the variational procedure, while insisting
that all equations are manifestly covariant. This means that when the electromagnetic forces in the physical
scenario do not on their own give the assumed motion (such as rigid motion), the application of the constraint
automatically gives rise to a further force which, although it may have some other physical origin, such
as a strong nuclear force, or the programmable device inside our imagined balloon, appears in the equations
as a modification of the force called `electromagnetic'. Having said that, it is appropriate to remark that what one
usually wants to know in practice is the total self-force for a given rigid body, i.e. the sum
$\ffF_{\rm self} + \ffF_{\rm P}$ in our notation, and this is what Nodvik has provided.

Finally, in view of the fact that for hyperbolic motion $\ddot{\ffU} = a_0^2 \ffU$
and using $\ffU^2 = -1$, we have $\ddot{\ffU} \cdot \ffU = -a_0^2$. By repeatedly
differentiating $\ddot{\ffU} = a_0^2 \ffU$ one finds
\be
a_0^{2n} = -  \ffU \cdot \ffU^{(2n)}
\ee
Hence another way to write
Eqn (\ref{fourself}) is
\be
\ffF_{\rm self} = \frac{2 e^2}{R} \dot{\ffU} \sum_{n=0}^\infty
\frac{-R^{2n} \ffU \cdot \ffU^{(2n)} }
{(2n-1)(2n+1)^2(2n+3)}.
\ee

\appendix{Series solution for the field}

We would like to find the derivatives of $\phi(x,0,0)$ with respect to $x$ at $x=1$. This amounts to finding the Taylor
expansion of $\phi(x,0,0)$ about $x=1$. To this end, introduce $h \equiv x-1$, then 
eqn. (\ref{phix00}) can be written
\be
\phi(x,0,0) = \phi_-(h) R^{-1} + \phi_+(h) R
\ee
where
\be
\phi_-(h) = \frac{1}{2(1+h)} + T(h), 
\;\;\;\;\;
\phi_+(h) = \frac{1}{h(2+h)} \left( \frac{-1}{2(1+h)} +T(h) \right)
\ee
with
\be
T(h) = \frac{h(1+h)(2+h)}{4( h(2+h) )^{3/2}} \tan^{-1} \left( \frac{2 \sqrt{h(2+h)} }{1 - h(2+h)} \right).
\ee
The first eleven terms in the Taylor expansions of $\phi_-$ and $\phi_+$ are
{\small
\[
\phi_- \simeq
1-\frac{h}{3}+\frac{2 h^2}{5}-\frac{46 h^3}{105}+\frac{29 h^4}{63}-\frac{547 h^5}{1155}+\frac{620
   h^6}{1287}-\frac{21932 h^7}{45045}+\frac{5959 h^8}{12155}-\frac{204739 h^9}{415701}+\frac{72614
   h^{10}}{146965}
\]
\[
\phi_+ \simeq \frac{1}{3}-\frac{7 h}{15}+\frac{18 h^2}{35}-\frac{166 h^3}{315}+\frac{365 h^4}{693}-\frac{523 h^5}{1001}+\frac{3332 h^6}{6435}-\frac{56204
   h^7}{109395}+\frac{117927 h^8}{230945}-\frac{164335 h^9}{323323}+\frac{342430 h^{10}}{676039}
\]
}

To find the self-force, we use the form (\ref{VN}), choosing a finite value for $N$ and finding the coefficients $a_n$
by matching the above Taylor series. To illustrate, we display the working for the case $N=3$. One finds the set of equations
\[
\left( \begin{array}{cccc}
1 & -1/2 & 1/3 & -1/4 \\
1 & -3/2 & 5/3 & -7/4 \\
0 & -3/2 & 10/3 & -21/4 \\
0 & -1/2 & 10/3 & -35/4 \end{array} \right)
\left( \begin{array}{c} a_0 \\ a_1 \\ a_2 \\ a_3 \end{array} \right)
= \frac{1}{R} \left( \begin{array}{c} 1 \\ -1/3 \\ 2/5 \\ -46/105 \end{array} \right)
+ R \left( \begin{array}{c} 1/3 \\ -7/15 \\ 18/35 \\ -166/315 \end{array} \right).
\]
The solution is
\[
a_0 = \frac{298}{105 R} +\frac{506 R}{315}, \;
a_1 = \frac{752}{105 R}  + \frac{544 R}{105},\;
a(2) = \frac{243}{35 R} + \frac{37 R}{7}, \;
a(3) = \frac{16}{7 R} + \frac{16 R}{9}.
\]
One can now find an approximation to $f_{\rm self}$ by using these values in (\ref{fselfsum}). One finds
\be
f_{\rm self} \simeq - \frac{2}{3 R} + \frac{2 R}{45} + \frac{2 R^3}{525} - \frac{466 R^5}{735} - \frac{32 R^7}{63}.   \label{fself3}
\ee
It is straightforward to perform a similar calculation at higher values of $N$. One thus obtains both an increasingly accurate
estimate of the self force, and also a power series approximation to the electric field inside the spherical shell. 
The terms in the power series for $V(h,\rho)$ are correct 
up to order $h^k \rho^p$ with $k+p=N$, and consequently the terms in the power series for the electric field
are correct up to order $h^k \rho^p$ with $k+p=N-1$. Therefore the series for $f_{\rm self}$ is
guaranteed to be correct up to order $R^{N-2}$ (not $N-1$ because each $a_n$ coefficient involves a $1/R$ term).

Comparing (\ref{fself3}) with the expression (\ref {fseries}) which was obtained using a higher value of $N$ so as to obtain
a higher order approximation to the potential, we observe that the solution with $N=3$ has correctly reproduced
the $O(1/R)$ term and the $O(R)$ term as expected, and it has also given the $O(R^3)$ term correctly. Is this just a coincidence?
Perhaps not, because a similar observation can be made about the $O(R)$ term in eqn (\ref{fsol2}). 
By examining the results for values of $N$ in the range $2$ to $20$, 
we found that the method at given $N$ yields the self force correctly up to order $R^{2N-3}$. In other words
it is surprisingly effective, converging more rapidly than was expected. However I have not been able
to discover why. This question might be explored by expanding the potential directly in terms of the $V_n$
functions, rather than going via the polynomial series. That is one possible direction for future work.

We will finish by illustrating this point in its simplest form, which is found when $N=2$.
The case $N=2$, calculated as above by matching the Taylor series on the $x$ axis, 
leads to the following approximation for the $x$-component of the electric field:
\be
E_x \simeq \left( -\frac{2}{3} + \frac{8}{15} h - \frac{212}{15} h^2 + \frac{104}{15} \rho^2 \right) R^{-1} + \frac{2}{15}R
\ee
whereas the correct power series expansion to this order is
\be
E_x \simeq \left( -\frac{2}{3} + \frac{8}{15} h - \frac{44}{105} h^2 + \frac{8}{105} \rho^2 \right) R^{-1} + \frac{2}{15}R .
\ee
The $N=2$ approximation does not give the terms of order $h^2/R$ and $\rho^2/R$ correctly. However, to find the self
force we evaluate the field at $\rho^2 + h^2 = R^2$ and then integrate over $h$ between $-R$ and $+R$. The calculation
using the correct expression for $E_x$ yields, for the coefficient of the term of order $R$ in $f_{\rm self}$:
\[
-\frac{44}{105} \times \frac{1}{3} + \frac{8}{105} \times \frac{2}{3} + \frac{2}{15} = \frac{2}{45}.
\]
and the calculation using the $N=2$ approximation to $E_x$ yields
\[
-\frac{212}{15} \times \frac{1}{3} + \frac{104}{15} \times \frac{2}{3} + \frac{2}{15} = \frac{2}{45}.
\]

\end{document}